\documentclass[
reprint,
showkeys,
longbibliography,
nofootinbib,
superscriptaddress,
aps,
notitlepage,pra
]{revtex4-1}
\usepackage[utf8]{inputenc}
\usepackage{bigints}
\usepackage{amsmath}
\usepackage{amssymb}
\usepackage{bm}
\usepackage{enumitem}
\usepackage[acronym]{glossaries}
\usepackage{graphicx}
\usepackage[colorlinks,unicode]{hyperref}
\usepackage[capitalise]{cleveref}  % must come after hyperref
\usepackage{mathtools}
\usepackage{multirow}
\usepackage{physics}
\usepackage{siunitx}
\usepackage[dvipsnames]{xcolor}
\usepackage{tcolorbox}

\DeclareSIUnit\week{week}

\makeatletter
\renewcommand\onecolumngrid{%
\do@columngrid{one}{\@ne}%
\def\set@footnotewidth{\onecolumngrid}%
\def\footnoterule{\kern-6pt\hrule width 1.5in\kern6pt}%
}
\makeatother

\newcommand\myshade{80}
\colorlet{mylinkcolor}{ForestGreen}
\colorlet{mycitecolor}{Red}
\colorlet{myurlcolor}{violet}

\hypersetup{
  linkcolor  = mylinkcolor!\myshade!black,
  citecolor  = mycitecolor!\myshade!black,
  urlcolor   = myurlcolor!\myshade!black,
  colorlinks = true
}

\newcommand{\calE}{\mathcal{E}}
\newcommand{\calR}{\mathcal{R}}
\definecolor{SeaGreen4}{rgb}{0.18, 0.55, 0.34}

\DeclareMathAlphabet\mathbfcal{OMS}{cmsy}{b}{n}
\DeclareSymbolFont{mathtx}{OML}{txmi}{m}{it}
\DeclareMathSymbol{v}{\mathalpha}{mathtx}{118}

\newcommand{\GRAPPA}{Gravitation Astroparticle Physics Amsterdam (GRAPPA),\\ University of Amsterdam, 1098 XH Amsterdam, The Netherlands}

\newcommand{\IFCA}{Instituto de F\'{i}sica de Cantabria (IFCA, UC-CSIC), Av. de Los Castros s/n, 39005 Santander, Spain}

\begin{document}

\title{On the survival of dark matter spikes: \\stellar and compact-object perturbations}

\author{Theophanes K. Karydas}
\email{t.karydas@uva.nl}
\affiliation{\GRAPPA}

\author{Francesca Scarcella}
\affiliation{\IFCA}

\author{Bradley J. Kavanagh}
\affiliation{\IFCA}

\author{Gianfranco Bertone}
\affiliation{\GRAPPA}

\begin{abstract}
Establishing realistic expectations for the dark matter (DM) distribution near a supermassive black hole (BH) is essential for assessing environmental imprints on gravitational wave (GW) signals. Using the Galactic Center as an observationally constrained case study, we investigate the evolution of DM density spikes under gravitational perturbations from the nuclear stellar and black hole populations surrounding the central BH. We find that scattering of DM particles by the nuclear star cluster depletes the DM distribution at radii $r \sim 10^{-1}~\mathrm{pc}$, far outside the region where the relevant GW signals are produced. At smaller radii, at $r \sim 10^{-3}~\mathrm{pc}$, the closest known stellar perturbers, S2 and S38, induce only negligible changes to the density profile. At still smaller radii, where no stellar perturbers are currently known, we assess the cumulative impact of past EMRIs by modelling successive mergers with stellar-mass BHs of mass $\sim10~\mathrm{M}_\odot$, in numbers consistent with the expected rate over $10~\mathrm{Gyr}$. We find that these events do not erase the central overdensity, reducing the density only to approximately $82 \%$ of its initial value at $r \lesssim 10^{-5}~\mathrm{pc}$. Our results indicate that, at least at the Galactic Center, DM overdensities around the central BH are expected to remain largely intact under stellar perturbations and plausible stellar-BH merger histories.
\end{abstract}

\maketitle

\section{Introduction}
Despite decades of sustained theoretical and experimental effort, the fundamental nature of dark matter (DM) remains elusive \cite{Bertone_20051,Feng:2010gw,Bertone_2018,Cirelli:2024ssz}. This has motivated the exploration of detection strategies that are
complementary to traditional laboratory and astrophysical searches. Future space-based gravitational-wave detectors
\cite{LISA:2024hlh,TianQin:2015yph,Hu:2017mde} will observe extreme
mass-ratio inspirals (EMRIs), opening a new window on both strong field
gravity and the astrophysical environments of compact binaries
\cite{Amaro_Seoane_2007,Gair_2013,Barack_2019,Bertone_2020,
Baumann_2022,Maselli_2022,miller2025gravitationalwaveprobesparticle}. In particular, EMRIs can probe the environment around these binaries~\cite{Cardoso:2019rou, LISA:2022kgy, CanevaSantoro:2023aol, Bertone:2024rxe, Zwick:2025wkt, Zwick:2025wkt, Cardoso_2022, spieksma2025blackholespectroscopyenvironments, Cole:2022yzw, Roy:2024rhe}, including DM overdensities, such as \emph{spikes} and \emph{mounds}~\cite{Gondolo_1999,Zhao_2005,Bertone:2005xz,Sadeghian_2013,Ferrer_2017,bertone2024darkmattermoundsrealistic,Ullio_2001,caiozzo2025darkmattermoundscollapse}. Such environments can modify the orbital evolution of the binary and
leave characteristic imprints on the emitted gravitational waveform~\cite{Eda_2013, Kavanagh_2020, Coogan_2022, Cole_2023, karydas2024sharpeningdarkmattersignature, Kavanagh:2024lgq, Speeney_2022, Becker_2022, mitra2025extrememassratioinspirals, Li_2022, Zhang_2024, Mukherjee_2024, Hannuksela_2020, Becker_2023, zhou2025intermediatemassratioinspiralsgeneraldynamical, Nichols_2023, Edwards_2020, Montalvo_2024, Vicente:2025gsg,karydas2025massspincoevolutionblack,karydas2025measuringneutronstarequation}, thereby providing
a complementary probe of dark matter.

The interaction of EMRIs with standard collisionless DM is mediated only by gravity. The leading environmental effects in this case arise from dynamical friction \cite{chandra1, chandra2, chandra3} and, for BH companions, from particle accretion~ \cite{Misner:1973prb,Traykova_2023,karydas2024sharpeningdarkmattersignature,karydas2025massspincoevolutionblack}. Both effects are sensitive to the details of the DM phase-space distribution \cite{Kavanagh_2020,karydas2024sharpeningdarkmattersignature,Becker_2022}. The detectability and interpretation of DM effects in EMRI waveforms therefore depend on the time-dependent DM distribution around the central massive black hole.

Several physical processes can affect such DM distributions.
It has long been recognized that stars can efficiently heat up and diffuse DM distributions on astrophysical timescales by gravitational scattering \cite{Merritt_2004,Bertone_2005,Gnedin_2004,Shapiro_2022,Balaji:2023hmy}. In addition, the orbital energy deposited by inspiraling BHs can further alter the DM environment \cite{Kavanagh_2020,Coogan_2022,Mukherjee:2023lzn,karydas2024sharpeningdarkmattersignature,Kavanagh:2024lgq}. Although individual minor mergers (with mass ratio $q \lesssim 10^{-4}$) induce only negligible changes in the density, supermassive black holes might experience $\mathcal{O}(100-1000)$ such events over their lifetime \cite{Amaro_Seoane_2011,Babak_2017}, raising the possibility of a non-negligible cumulative impact.

In this work, we show that the DM spike around Sgr A* at the Galactic Center is remarkably robust against gravitational interactions with stars and stellar BHs. Including scattering by the nuclear star cluster and the perturbing effect of stars in the S-cluster at the Galactic center, we find that the DM density is essentially preserved in the innermost region relevant for GW dephasing, $r\sim10^{-5}~\mathrm{pc}$. To account for the cumulative impact of past EMRIs, we also model successive mergers of a population of $\sim10~\mathrm{M}_\odot$ black holes over $10~\mathrm{Gyr}$, finding that they reduce the density in this region only modestly, by $\sim18 \%$.

\section{Depletion from the nuclear star cluster} \label{sec:nsc}
A dense nuclear star cluster (NSC) surrounds the Galactic Center, extending out to radii of order $\sim 5~\mathrm{pc}$~\cite{Neumayer_2020,Sch_del_2009,refId0,Genzel_2010}. Beyond this radius, the stellar density falls steeply as $\rho_\star \propto r^{-3}$~\cite{Sch_del_2014}, whereas inside $\sim 5~\mathrm{pc}$ the NSC hosts a much shallower cusp. Within this radius, the old, main-sequence stellar population has been found to be consistent with a cusp $\rho_\star \propto r^{-1.4}$ for $r \gtrsim 1" \approx 0.04~\mathrm{pc}$~\cite{2018A&A...609A..26G}. In contrast, the observed distribution of giant stars remains significantly flatter within $\sim 0.1~\mathrm{pc}$ of Sgr A*, possibly owing to stellar collisions~\cite{Rose_2024}, envelope stripping by a past stellar disk \cite{Amaro-Seoane_2014}, or other dynamical effects.

The NSC is unlikely to be fully relaxed, as the two-body relaxation time at parsec scales is of order, or longer than, its age \cite{2013degn.book.....M}. This is corroborated by its kinematics, which show evidence for tangential anisotropy. The velocity anisotropy is commonly characterized by $\beta \equiv 1 - {\sigma_t^2}/(2\sigma_r^2)$, where $\sigma_{t,r}$ are the tangential or radial velocity dispersions \cite{binney}. Within $0.5~\mathrm{pc}$, the NSC has been found to be mildly tangentially anisotropic, with $\beta \approx -0.3$ \cite{Feldmeier_Krause_2016}.

This stellar system fully encompasses the dark matter (DM) spike expected to form around the central supermassive black hole, extending out to $\sim 0.34~\mathrm{pc}$ \cite{Bertone_2005}. Stars in the NSC gravitationally scatter individual DM particles, heating the DM distribution toward kinetic energy equipartition with the stellar component \cite{2013degn.book.....M}. As a consequence, the central spike is gradually depleted and the DM density is substantially reduced \cite{Merritt_2004,Bertone_2005,Shapiro_2022}. This has been demonstrated with isotropized orbit-averaged Fokker-Planck equations and fluid conduction models (see refs above). However, given the modeling assumptions of frequent and independent encounters in these two approaches, the inferred density evolution is physically well-motivated in regions where the stellar population is sufficiently abundant over the timescales. Thus, comparable depletion is not implied, by itself, in the innermost region relevant for influencing the gravitational-wave signals of EMRIs where the mean-field representation of the NSC breaks down.

As a concrete example, within $0.5~\mathrm{pc}$ we adopt the fiducial density model for the NSC of \cite{Bertone_2005,2018A&A...609A..26G,Genzel_2003} (and, for completeness, the associated DM spike profile),
\begin{align}
    \rho_\mathrm{DM}(r) &= 40 \left(r/\mathrm{pc}\right)^{-7/3}  &M_\odot/\mathrm{pc}^3\,,\\
    \rho_\mathrm{s}(r) &= 3 \times 10^5 \left(r/\mathrm{pc}\right)^{-1.4} & M_\odot/\mathrm{pc}^3\,,
\end{align}
which is observationally constrained down to $r \simeq 0.04~\mathrm{pc}$ \cite{2018A&A...609A..26G,Sch_del_2020}. Below this threshold, the assumptions underlying the Fokker–Planck equation are no longer valid, and inward extrapolation of the profile would lead to an overestimation of the depletion, which we quantify in the following section using an alternative approach. Within the Fokker–Planck framework, we therefore model both the stellar density above the threshold and the inward contribution from stars whose apocenters lie outside the threshold but whose orbits penetrate it.

The mass in stars whose orbits cross this threshold would form a shallow core with $\rho \propto r^{-(0.5+\beta)}$ \cite{An_2006}. For tangentially anisotropic systems such as the NSC \linebreak ($\beta < 0$), high-eccentricity orbits are suppressed, leading to a shallower profile. Taking $\beta =-0.3$ in this region \cite{Feldmeier_Krause_2016}, we can approximate the enclosed mass to be $\sim 10^3~M_\odot$. To remain conservative, however, we model the stellar distribution as isotropic ($\beta = 0$), impose an inner cutoff at $0.04~\mathrm{pc}$, and for $r < 0.04~\mathrm{pc}$ we model the density of those stars whose apocenters are outside this region by the steeper-than-expected $\rho \propto r^{-0.5}$ density profile.

Given these assumptions for the NSC density profile, we proceed to solve the anisotropic, orbit-averaged Fokker–Planck equation (see \cite{2017ApJ...848...10V} and references therein). In this framework, we describe the secular evolution of DM particle energy and, crucially, angular momentum due to scattering off the NSC, and its implications for the phase-space and density evolution of the DM spike.

\subsection{Fokker–Planck and Stochastic Description of DM Scattering}
We consider a spherically symmetric system of DM particles orbiting a supermassive BH of mass $M$ far from its strong-field gravitational effects and where their average velocity is non-relativistic. Their Keplerian motion is fixed by two isolating integrals of motion: the energy $\calE$ and angular momentum $h$ per unit mass $m_\chi$, or alternatively the normalized angular momentum \linebreak $\mathcal{R}=2\calE h^2/(GM)^2 \in (0, 1]$. The density of particles in phase space is given by 
\begin{equation}
    N(\calE, \mathcal{R})\,\mathrm{d}\calE\,\mathrm{d}\mathcal{R}=g(\calE,\mathcal{R}) f(\calE, \mathcal{R})\,\mathrm{d}\calE\,\mathrm{d}\mathcal{R}\,,
\end{equation}
where $f(\mathcal{E},\mathcal{R})$ is the distribution function, and given a potential $\Psi$,
\begin{align}
    g(\calE, \mathcal{R} \, | \, \Psi) &\equiv \int \left|\frac{\partial\left(v_r, v_\theta, v_\phi|\Psi \right)}{\partial\left(\calE, \mathcal{R}, h_z\right)}\right| \mathrm{d}^3 \mathbf{r} \,\mathrm{d}h_z\,,
\end{align}
is the density of states \cite{2013degn.book.....M}; for a point-mass potential we find $g(\calE, \mathcal{R})=\sqrt{2}\pi^3 G^3 M^3 \calE^{-5/2}$. We obtain the distribution function via Eddington inversion \cite{binney}, assuming initial isotropy, an approximation consistent with the modest radial bias of adiabatically grown DM spikes which have $\beta = 0.3$ \cite{Gondolo_1999}.

Particles undergo stochastic gravitational encounters with stars and receive phase-space kicks: $\Delta \calE$ and $\Delta \calR$ that perturb their orbits. The secular evolution of the DM particle density in phase-space is governed by the two-dimensional Fokker-Planck equation given in \textit{flux-conservation form} below \cite{2013degn.book.....M}:
\begin{equation}
    \frac{\partial N}{\partial t} = -\frac{\partial F_\calE}{\partial \calE} - \frac{\partial F_\mathcal{R}}{\partial \mathcal{R}}\,, \label{eq:fokker-planck}
\end{equation}
with the fluxes given by,
\begin{align}
    -F_\calE &= D_\calE N +D_{\calE\calE} \frac{\partial N}{\partial \calE} +D_{\calE\mathcal{R}} \frac{\partial N}{\partial \mathcal{R}} \,,\\
    -F_\mathcal{R} &= D_\mathcal{R} N +D_{\mathcal{R}\mathcal{R}} \frac{\partial N}{\partial \mathcal{R}} +D_{\calE\mathcal{R}} \frac{\partial N}{\partial \calE} \,,
\end{align}
where $D_i$, $D_{ij}$ are the first- and second-order diffusion coefficients which we derive in Appendix~\ref{app:coeffs} generalizing \citet{2013degn.book.....M} to account for the geometry of the relative encounters between stars and DM particles. These coefficients depend on the average rate of kicks, e.g. $\langle \Delta \calE \rangle_t$ or $\langle \Delta \calE \Delta \calR \rangle_t$.

\begin{figure}[ht!]
    \centering
    \hspace*{-0.08\columnwidth}
    \includegraphics[width=0.95\linewidth]{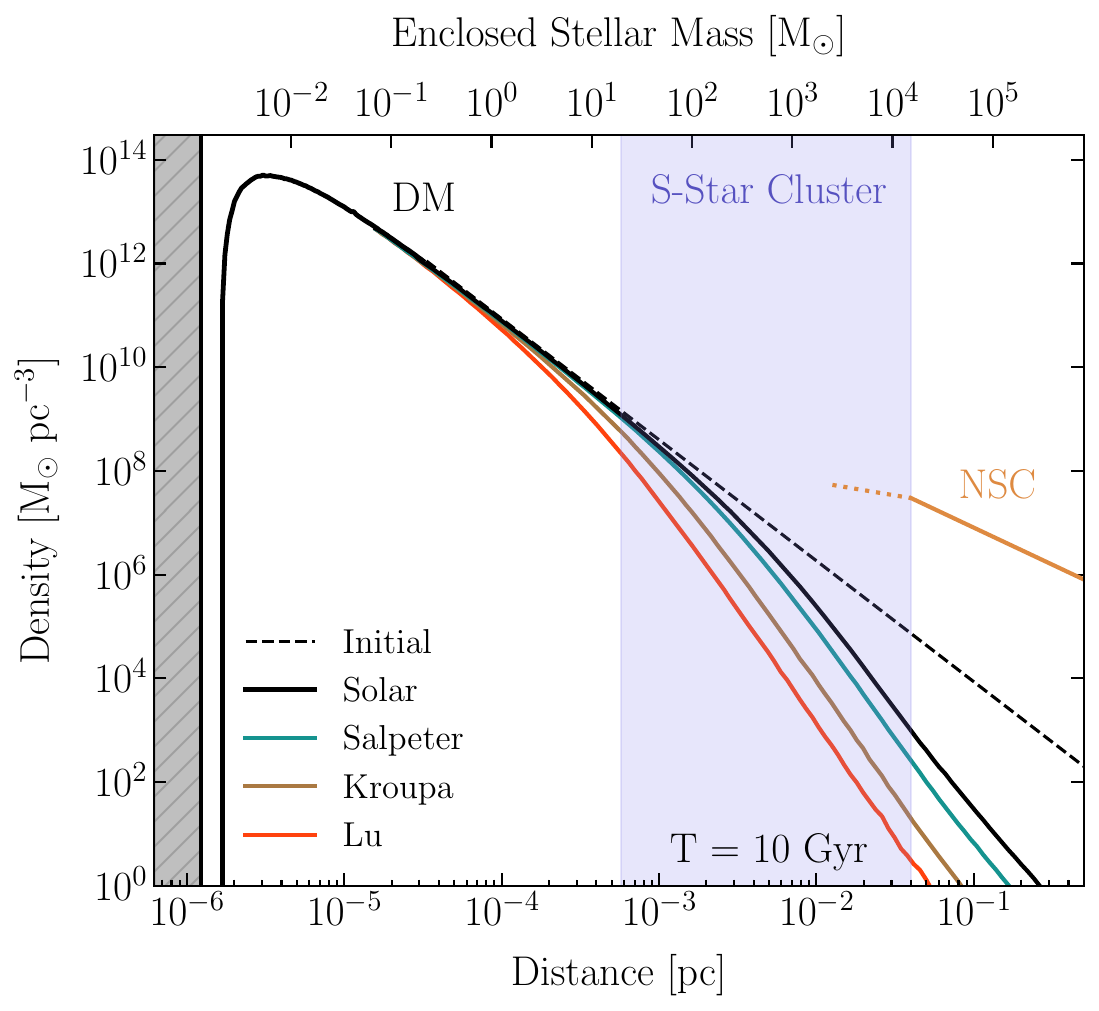}
    \caption{\textbf{Depletion of the dark matter density due to scattering with the nuclear star cluster under different stellar mass functions.} The purple region indicates the extent of the S-star cluster, defined to encompass the orbit of the S2 star and up to $0.04~\mathrm{pc}$ \cite{Sch_del_2020}. For reference, we show the mass density profile of stars whose apocenters are contained entirely outside $0.04~\mathrm{pc}$ but truncate the core for visual clarity. On the top axis, we indicate the enclosed stellar mass assuming the power-law density profile extends smoothly into the inner regions to illustrate their sparsity.\label{fig:NSC}}
\end{figure}

Equivalently to \cref{eq:fokker-planck}, we will instead formulate the evolution of the system in terms of an Itô stochastic differential equation \cite{1360855570504742656} for individual particle trajectories in $(\calE,\mathcal{R})$ space. In this representation, the system is advanced in time by evolving the coordinate position of pseudo-particles in phase-space according to,
\begin{align}
    \mathrm{d}\calE &= \langle \Delta \calE \rangle_t \, \mathrm{d}t + \sqrt{D_{\calE\calE}}\,dW_\calE\,,\label{eq:diffusion_equations}\\
    \mathrm{d}\mathcal{R} &= \langle \Delta \mathcal{R} \rangle_t \, \mathrm{d}t + \!\sqrt{D_{\mathcal{R}\mathcal{R}} \!-\frac{ D^2_{\calE\mathcal{R}}}{D_{\calE\calE}}}\,\mathrm{d}W_\mathcal{R} +\frac{D_{\calE \mathcal{R}}}{\sqrt{D_{\calE\calE}}} \,\mathrm{d}W_\calE\,, \nonumber
\end{align}
where $\mathrm{d}W_\calE$ and $\mathrm{d}W_\mathcal{R}$ are independent Wiener increments with $\langle \mathrm{d}W_i \rangle = 0$ and $\langle \mathrm{d}W_i \mathrm{d}W_j \rangle = 2 \,\delta_{ij}\,\mathrm{d}t$. The initial distribution of the pseudo-particles is obtained by the isotropic probability density function $N(\calE,\mathcal{R}, t= 0)$, and the system is simulated in time stochastically under \cref{eq:diffusion_equations}. The phase-space density at a given moment in time, $N(\calE,\mathcal{R},t)$, is then reconstructed from the ensemble and used to calculate the density profile
\begin{align}
    \rho(r,t) &= m_\chi \int f(\calE, \mathcal{R}, t)\,\mathrm{d}^3 \mathbf{v}\\
    &= m_\chi \pi \sqrt{2} \frac{GM}{r} \bigintssss \frac{N(\calE, \mathcal{R}, t) \,\mathrm{d}\calE\,\mathrm{d}\mathcal{R}}{g(\calE,\mathcal{R})\,\sqrt{\calE}\sqrt{\frac{4\calE\left(\Psi-\calE\right)}{\Psi^2}-\mathcal{R}}}\,, \nonumber
\end{align}
where $\Psi = GM/r$. In our simulation, particles are allowed to become unbound from the system $(\calE < 0)$ or to be captured by the black hole through the weak-field loss-cone condition, $\mathcal{R} \leq 32 \, \calE/c^2$ \cite[see Eq. 3.16]{Sadeghian_2013}. To maintain statistical precision as the number of particles decreases, we periodically replenish the lost points by randomly splitting bound particles and readjusting the weights of each copy when estimating $N(\calE, \mathcal{R})$.

In \cref{fig:NSC} we show in solid lines the DM density profile after evolving for $10~\mathrm{Gyr}$ in the presence of the NSC and for reference also show the NSC density profile in light orange). Although the spike is locally depleted by the NSC, the density rapidly returns to its initial profile in the innermost region. We furthermore observe that the system develops a ``broken'' power–law structure: the initial inner spike is preserved, and beyond a break radius (which shrinks as time progresses), the profile transitions to a much steeper slope. Following arguments based on the \emph{isotropized} Fokker–Planck equation, the DM distribution around a BH is expected to relax toward a steady-state cusp $\rho_{\rm DM} \propto r^{-3/2}$ \cite{2013degn.book.....M}. In contrast, our model that treats changes in $\calE$ and $h$ produces a significantly steeper inner profile. DM particles are efficiently diffused in the region where the stellar population is dense, but remain largely unaffected inside the transition radius. The steeper power law then arises as the depleted outer profile rises to match the essentially unperturbed inner spike. Finally, we note that at larger radii, $r \sim 0.5\,\mathrm{pc}$, the gravitational potential is dominated by the BH and the depletion is expected to slow down; this regime lies outside the domain of validity of our model.

\subsection{Dependence on the Stellar Mass Spectrum}
We now investigate the dependence of the DM distribution on the stellar mass spectrum. The scattering strength and the relative abundance of stars depend on the mass distribution within the population, and variations can therefore have a significant impact on the heating evolution of the DM spike \cite{Bertone_2005, Merritt_2004}. We model the mass spectrum using the stellar initial mass function (IMF), which parametrizes the relative distribution of stellar masses at birth, and assume it is spatially uniform and constant in time, while noting that more massive objects naturally relax toward the cluster center. In this study, we adopt a set of IMFs motivated by observations of the Milky Way NSC and by theoretical studies; for comparison we consider the standard IMF's proposed by Salpeter \cite{salpeter} and Kroupa \cite{kroupa}, as well as the one proposed by Lu \cite{Lu_2013} based on observations of the cluster, and for reference, a single mass population of solar mass objects.

For each IMF, we re-simulate the system for $10~\mathrm{Gyr}$ while holding all other parameters fixed. In \cref{fig:NSC} we show the evolution of the corresponding DM density profile under each IMF assumption in different color. We find that increasingly top-heavy IMFs lead to more pronounced depletion of the density profile above $10^{-3}~\mathrm{pc}$. In this region, the profiles follow a simple scaling relation between IMFs: the ratio of any two curves is well described by a factor $\left( \langle m^2 \rangle / \langle m \rangle \right)^2 > \langle m \rangle^2$, where $\langle m^k \rangle$ denotes the $k$-th mass moment of the IMF. This scaling reflects the fact that, in our diffusion treatment, the typical velocity kicks induced by a compact object of mass $m_s$ scale as $\langle m_s^2\rangle$, so that more massive black holes impart disproportionately stronger kicks, which more than compensate for their lower number density. As a result, top-heavy IMFs, with larger $\langle m^2 \rangle / \langle m \rangle$, produce stronger diffusion and hence a more strongly depleted outer profile.

\section{Depletion from the S-star cluster}

Further inward, the NSC becomes increasingly sparse, eventually transitioning to the S-star cluster. In this regime, the assumptions underlying the Fokker-Planck description break down, as diffusion is driven by a small number of discrete perturbers rather than a smooth background described by a mean-field distribution function. The most relevant perturbers here are young S-stars with ages $\sim \mathrm{Myr}$, so any heating they induce cannot act over a Hubble time. To quantify the depletion induced by an $\mathcal{O}(1)$ population of young stars residing entirely within $0.04~\mathrm{pc}$, and to assess their impact on the dark-matter spike, we employ the ``feedback'' framework of Refs.~\cite{HaloFeedback,Kavanagh_2020,karydas2024sharpeningdarkmattersignature}. This formalism captures dark-matter ``heating'' due to scattering off individual objects, including strong kicks, by working at the master-equation level~\cite{2013degn.book.....M}. In doing so, it retains the full jump statistics rather than relying on truncated Fokker–Planck diffusion moments, which are valid only in the small-kick limit.

For concreteness, we focus on the S2 star \cite{S2_star}, whose short-period and tightly bound orbit has one of the smallest semimajor axes among the S-stars and therefore the largest potential to perturb the deepest regions of the DM distribution. Its orbit is well measured from long-term monitoring ($e \approx 0.88$, $p\approx 118.4~\mathrm{AU}$, $m\approx14~\mathrm{M}_\odot$), and its formation is inferred to have happened about $6~\mathrm{Myr}$ ago \cite{Habibi_2017}. So any heating it induces can only operate over its relatively short lifetime.

During each orbital period, S2 gravitationally perturbs the surrounding DM, and gradually reshapes the density locally \cite{Kavanagh_2020}. The rate of change of the (isotropized) DM spike's distribution function due to gravitational scattering off a single star on a prescribed orbit, can be written instantaneously as
\begin{equation} \label{eq:feedback}
    \frac{\mathrm{d}f}{\mathrm{d}t} = \frac{1}{T_s}\left( \Delta_+ f - f\int P_\calE (\Delta \calE)\,\mathrm{d}\Delta\calE \right)\,,
\end{equation}
where $T_s$ is the star's orbital period taken to be Keplerian, $P_\calE(\Delta\calE)$, given in given in \cite{Kavanagh_2020,karydas2024sharpeningdarkmattersignature}, is the probability that a particle with specific energy $\calE$ scatters off the star and is boosted by the energy kick $\Delta\calE$; therefore, the second term represents the total scattering probability at the state $\calE$. Conversely,
\begin{equation}
    \Delta_+f = \!\int\! \left( \frac{\calE}{\calE -\Delta\calE} \right)^{5/2} \!\!\!f(\calE-\Delta\calE) P_{\calE-\Delta\calE}(\Delta\calE)\,\mathrm{d}\Delta\calE,
\end{equation}
represents the flow of particles to a different energy level. After $df/dt$ is calculated instantaneously, the total rate is averaged over the geometrical torus drawn by the star \cite{karydas2024sharpeningdarkmattersignature}.

In practice, we evolve the DM distribution function by solving the master-equation system using \texttt{HaloFeedback}~\cite{HaloFeedback,Kavanagh_2020,karydas2024sharpeningdarkmattersignature}. Following Ref.~\cite{Kavanagh:2024lgq}, we assume that the maximum impact parameter for which particles can be efficiently scattered by the star scales with its instantaneous separation, $r$, and we adopt $b_\mathrm{max} = 0.3\,r$.

These equations capture the strong, two-body deflection of particles off the star; however, this description is only accurate for a subset of particles undergoing close encounters with the perturber, where their trajectories are significantly perturbed. The dynamics for particles on wider orbits cannot be reduced to isolated two-body scattering, since they simultaneously feel the gravitational influence of both BH and star over extended timescales. In this regime, the evolution is inherently a more complicated three-body problem. As shown in \cite{Kavanagh:2024lgq}, this dynamical effect can be captured by modeling the particle's energy drift induced by the modulation of the orbiting star's gravitational potential known as the \textit{stirring}-effect. When the orbital evolution timescale is longer than the depletion timescales, this effect tends to bring particles in from larger radii, and it has been shown that by including it, the largest overall suppression in density is smaller as long as the depletion is small (cf. \cite[Fig. 7]{Kavanagh:2024lgq}). Because our goal is to obtain a conservative estimate of the star's impact on the innermost DM density profile, we neglect this contribution in order to simplify the analysis in this section.

To obtain a conservative estimate for the heating induced by the S2-star, we maximize its impact by considering the following scenario: we assume that S2 has remained on a fixed orbit since its formation. Evolving the DM density under this assumption, we find that the resulting maximum depletion is negligible over the lifetime of the star: the maximum fractional change relative to the initial density $\rho_0$ is well approximated by
\begin{equation}
    \frac{\Delta \rho}{\rho_0} \approx 2.2 \times 10^{-5} \left( \frac{t}{\mathrm{Myr}} \right).
\end{equation}
Even for $t \sim 10~\mathrm{Myr}$, it yields only $\Delta \rho / \rho_0 \sim 10^{-4}$. This confirms that stellar heating by S2 is too weak to significantly perturb the innermost DM spike. The rest of the young stars in the S2 cluster have a similarly negligible impact on the DM distribution. 

Finally, we must quantify the effect of the older giants which operate over Gyr timescales instead. Following the same procedure as above, we study the impact of S38 ($e\approx 0.82$, $p\approx 209$, $m\approx 5.97~\mathrm{M}_\odot$) \cite{2019ApJ...872L..15H} on the DM distribution. Our estimate of the depletion is again conservative, as we fix the stellar orbit and model the star as a point-mass. The latter assumption maximizes its perturbative potential given S38's large radius ($R\approx 186~R_\odot$) which suppresses dynamical friction. Again, we find a very modest depletion over the relevant timescales which can be fitted by the formula
\begin{equation}
    \frac{\Delta \rho}{\rho_0} \approx 3.8 \times 10^{-3} \left( \frac{t}{\mathrm{Gyr}} \right).
\end{equation}
Similarly, a population of BHs and other compact objects predicted by mass-segregation models \cite{2024ApJ...961..232Z} in this regime would induce a small impact.

\section{Depletion from past EMRIs} \label{sec:emris}
Over the past $\sim 10~\mathrm{Gyr}$, a population of stellar-mass BHs is expected to have migrated from the NSC deep into the Galactic Center, and subsequently merged with the central SMBH. In doing so, these inspirals each probe the innermost regions of the DM spike and have the potential to perturb its density profile. For extreme mass ratios with $m/M \lesssim 10^{-4}$, the effect of an individual merger on the DM profile is small \cite{Kavanagh_2020,Coogan_2022}, since DM particles are largely scattered back onto nearby orbits as the merger progresses. Nevertheless, these studies have found a small residual suppression of the spike after each event. The cumulative impact of many such mergers over a Hubble time has not yet been systematically investigated; in this section, we set out to quantify this effect.

Each merger will induce a rate of change in the distribution function given by \cref{eq:feedback}. In addition, another contribution must be included to account for the accretion of DM particles onto the event horizon of the inspiralling companion:
\begin{equation}
    \frac{\mathrm{d}f_\mathrm{acc}}{\mathrm{d}t} = -\frac{p_\mathrm{acc}(\calE)}{T} f\,,
\end{equation}
where $p_\mathrm{acc}(\calE)$, given in \cite{karydas2024sharpeningdarkmattersignature}, describes the probability of particles with energy $\calE$ being accreted by the companion, and $T_s \to T$ is the orbital period of the companion.

\begin{figure}[t]
    \centering
    \hspace*{-0.08\columnwidth}
    \includegraphics[width=0.95\linewidth]{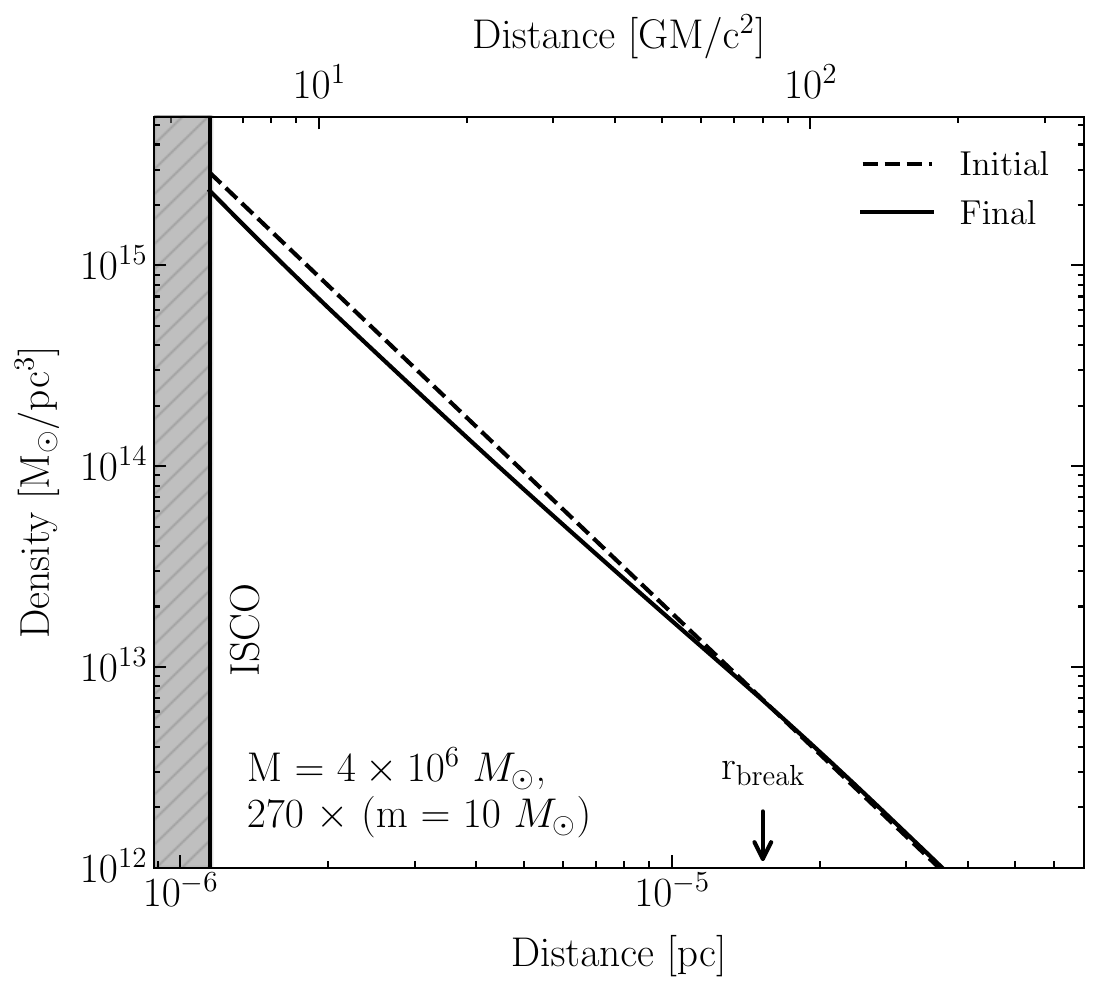}
    \caption{\textbf{Depletion of the dark matter density due to past mergers with a BH population over $\mathbf{10}$~Gyr.} The arrow indicates $r_\mathrm{break}\approx 1.53\times 10^{-5}~\mathrm{pc}$, as derived in Ref.~\cite{Coogan_2022}, which marks the radius within which the gravitational-wave emission timescale becomes shorter than the depletion timescale. Assuming 270 past events each with companion mass $m=10~\mathrm{M}_\odot$.\label{fig:consecutive_mergers}}
\end{figure}

The inspiral of each BH within the DM spike is driven by the emission of GWs \cite{Misner:1973prb}, as well as by dynamical friction and accretion arising from the spike. A self-consistent treatment therefore requires evolving both the distribution function and the orbital parameters of the inspiraling BH. In particular, the equations of motion must include the additional forces along the direction of motion $\mathbf{u}$,
\begin{align}
    F_\mathrm{df} &= -4 \pi G^2 m^2 \frac{\rho}{u^2} \mathcal{C}_\mathrm{df} \,,\\
    F_\mathrm{acc} &= -16 \pi G^2 m^2 \frac{u^2+c^2}{c^4} \rho\, \mathcal{C}_\mathrm{acc} \,,
\end{align}
as well as the increase in BH mass due to accretion:
\begin{equation}
    \frac{\mathrm{d}m}{\mathrm{d}t} = 16\pi G^2m^2 \frac{u^2+c^2}{u\,c^4} \rho \, \mathcal{C}_\mathrm{m}\,.
\end{equation}
Here the three coefficients $\mathcal{C}_j = \mathcal{C}_j(r, u, t)$,  which are defined in Ref.~\cite{karydas2024sharpeningdarkmattersignature}, as well as the time-dependent density $\rho(r) = \rho(r,t)$, encode the inspiral's dependence on the phase-space distribution of the spike. 

While the three-body stirring effect can be neglected when modelling a static companion in order to obtain a conservative estimate, it is not a priori clear whether this remains justified in an evolving system. In Appendix~\ref{app:stirring}, we present simulations demonstrating that this contribution can indeed be safely neglected in the innermost regions, even for inspirals with mass ratios as large as $q = 10^{-3}$. Based on this result, and given that our focus is on the inner region of the DM spike, we also neglect the stirring effect in the following analysis for the sake of simplicity, and use \texttt{HaloFeedback} to model the interactions.

\begin{figure*}[ht!]
    \centering
    \hspace*{-0.08\columnwidth}
    \includegraphics[width=0.93\linewidth]{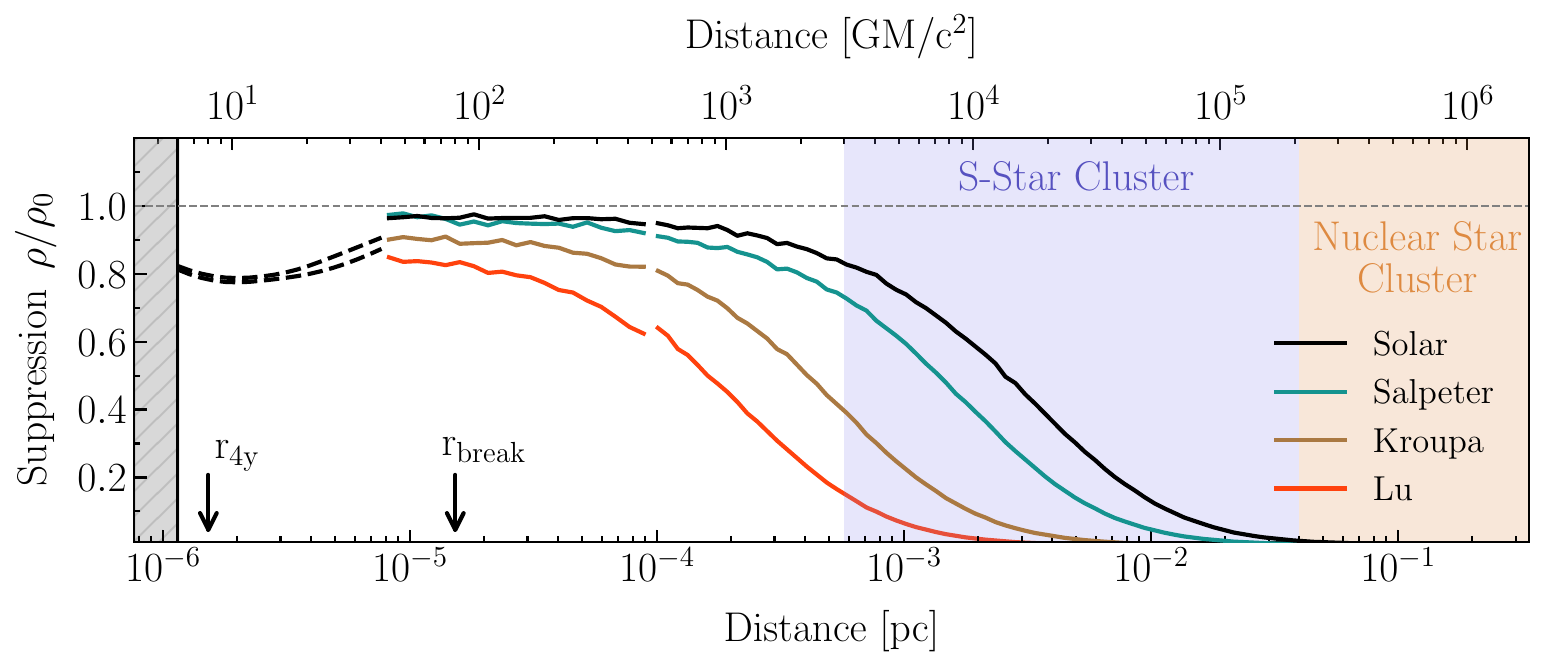}
    \caption{\textbf{Total suppression in the dark matter density of the spike in different regions.} The solid lines show the density suppression induced by the stellar population under different stellar mass function assumptions. Due to resolution limitations, the profiles are discontinuous at $10^{-4}~\mathrm{pc}$, where results from additional simulations are used. The dashed lines illustrate the suppression for two  different eccentricity realizations of the primary's EMRI history. The arrow marks the break radius where the depletion and inspiral timescales are equal. Finally, the purple region indicates the extent of the S-star cluster, defined to encompass the orbit of the S2 star up to $0.04\,\mathrm{pc}$. Within this, the S2 star depletes the profile by $\Delta\rho/\rho_0 \sim 10^{-4}$ and the older giant S38 by $\sim 10^{-3}$, effects too small to be visually discernible in the plot. The arrows shows the breaking radius \cite{Coogan_2022} discussed in the text, and $r_{4y}\approx8GM/c^2$ the separation with which the $10~\mathrm{M}_\odot$ companion merges with the primary after 4 years.
    \label{fig:relative}}
\end{figure*}

Each merger is initialized with a semi-major axis $a_0 = 720~\mathrm{GM/c^2}\approx 1.4\times 10^{-4}~\mathrm{pc}$, and an eccentricity drawn from the eccentricity distribution reported by \citet{Hopman_2005} at a reference $a = 17~\mathrm{GM/c^2}$. The eccentricity is then conservatively evolved backwards to $a_0$ assuming evolution driven solely by GW emission, which we verify to be accurate to better than $1\%$. As the BH companion mass distribution is largely unconstrained, we take the BHs to be of $10~\mathrm{M}_\odot$ and following \citet{Babak_2017} estimate that a $M=4\times10^6~\mathrm{M}_\odot$ BH, such as Sgr A*, would have undergone approximately $\sim$270 such events over $10~\mathrm{Gyr}$. Each inspiral lasts approximately $10^4-10^5~\mathrm{yrs}$, so we model the events consecutively.

In \cref{fig:consecutive_mergers} we show the cumulative effect of past mergers on the DM density profile. We observe that a well-defined region emerges, where the spike's density is decreased by $\sim 17\%$ while the logarithmic slope increases modestly towards the center. Following \citet{Coogan_2022}, we identify the onset of this region with the break radius, $r_\mathrm{break}$, defined as the radius within which the inspiral timescale due to gravitational-wave emission becomes shorter than the timescale for evolving the distribution function. Inside $r_\mathrm{break}$, the binary cannot efficiently replenish the DM hole carved by earlier parts of the inspiral leaving an imprint that accumulates with each merger. We find that the maximum suppression in density after $N$ such mergers can be approximated well by the following expression
\begin{equation}
    \Delta \rho/\rho_0 \approx 7\times 10^{-4} \, N\,.
\end{equation}

Finally, the total density suppression across the DM spike is shown in \cref{fig:relative}. Since the effect of the S2 and S38 stars are negligible compared to the depletion induced by the NSC at the same scale, it is omitted from the figure.

\section{Discussion \& Conclusions}

We do not specify the formation history of the BH, which remains unknown for the MW. However, any modification of the DM spike arising from nonadiabaticity during its formation would be erased during the accretion stage of the BH. A mass increase by a factor of $\sim 5$ is sufficient for the BH to adiabatically contract the surrounding DM and recover the spike profile adopted here~\cite{caiozzo2025darkmattermoundscollapse}.

While we have focused on the impact of minor mergers on the DM spike, we have not addressed the consequences of the expected $\mathcal{O}(1)$ major merger events during the lifetime of the supermassive BH. For mass ratios $q \lesssim 0.1$, mergers are not expected to significantly disrupt the DM spike \cite{merritt_2002} and will generally do so in the outermost regions. Moreover, a merger would funnel gas toward the central region, fueling BH growth and potentially increasing the DM density through subsequent adiabatic contraction which would partially replenish the DM density profile \cite{caiozzo2025darkmattermoundscollapse}. We defer a detailed numerical investigation of major mergers and their net impact on the spike to future work.

While our analysis has focused on the Galactic Center and a merger history appropriate to it, an alternative and potentially interesting scenario arises for mergers occurring in active galactic nuclei (AGNs). In these environments, binaries are expected to form in-situ within the accretion disk and to circularize efficiently. In the limit where all past inspirals have been quasi-circular, we find a substantially depleted region in the DM distribution within $r_\mathrm{break}$ even after $100$ mergers with $10~M_\odot$ BHs. The density in the region is suppressed by a little more than an order of magnitude, and the logarithmic slope increases towards the center becoming approximately $\gamma \sim 3.5$. This toy-model result suggests that the observation of an unusually steep DM spike could provide insight into the astrophysical environment of the central supermassive BH. A detailed exploration of this scenario, however, lies beyond the scope of the present work.

In this work, we assume a radius-independent mass probability density for the NSC, anchored to observational constraints in the Galactic Center. While adequate for our purposes, a more general treatment would require specifying an initial mass function and consistently evolving stellar lifetimes, remnant formation, and accretion, alongside mass segregation of heavier objects toward the center. In regimes where a small number of massive bodies dominate, the continuum approximation breaks down as the Fokker–Planck approach is no longer valid for an $\mathcal{O}(1)$ population. In such cases, direct N-body simulations may provide a more appropriate description to capture discreteness.

During the preparation of this work, Ref.~\cite{sharpe2026depletioncollisionlessdarkmatter} appeared and addressed a qualitatively similar question, but reported substantially stronger DM depletion due to past EMRIs than found here. Reference~\cite{sharpe2026depletioncollisionlessdarkmatter} performed scattering experiments to estimate the ejection probability $ p_\mathrm{ej}$ of DM particles through encounters with the EMRI over a single cycle. Assuming that this average probability does not vary from one cycle to the next, and that the particles otherwise remain in the same states, one finds $\dot{\rho} \propto p_\mathrm{ej} \rho$, leading to an apparent rapid exponential depletion.

In a more realistic description, however, particles that are not completely ejected in a single cycle do not remain in their original state in phase-space. Instead, they are re-distributed in phase-space, altering the distribution function through both ejection and state evolution\footnote{This redistribution also partially replenishes the outer parts of the spike, which may have been previously depleted during the inspiral.}. As a result, the survival probability after $N$ binary cycles is not given by repeated application of the initial survival probability $p_\mathrm{surv} = \left(1-p_\mathrm{ej,0}\right)^N$ which would imply exponential depletion. Rather, the ejection probability changes from cycle to cycle as the DM particle occupies different states, and the survival probability must be determined by tracking the sequence of possible new states the DM particle has occupied prior to ejection, $p_\mathrm{surv} = \left(1-p_\mathrm{ej,0}\right)\left(1-p_\mathrm{ej,1}\right)\left(1-p_\mathrm{ej,2}\right)\ldots$, and so on. 
The depletion is therefore determined by the full evolution of the distribution function. Non-ejecting encounters tend to redistribute particles into orbits which are harder to deplete, but which still contribute to the inner density. The ejection probability therefore decreases with successive cycles, which ultimately leads to the smaller-than-exponential depletion observed here.
This departure from exponential behaviour is clearly observed in the depletion rates obtained with \texttt{HaloFeedback}.

This slower-than-exponential behaviour has also been observed in $N$-body simulations using the \texttt{NbodyIMRI} code~\cite{Kavanagh:2024lgq,Scarcella:2025dsh,LongTermDepletion}. These Newtonian simulations allow us to evolve the full DM distribution over 1000s of IMRI/EMRI (see e.g.~Ref.~\cite{Kavanagh:2024lgq} for the $q = 10^{-2}$ case).  In these simulations, the DM density at the orbital radius of the binary shows an initial transient period of exponential depletion but then progresses to a lower depletion rate, as expected if the typical depletion rate is not constant but instead decreasing with time. More specifically, the depletion rate transitions to a power-law scaling, as discussed in \cite{LongTermDepletion}.

In conclusion, we have shown that the expected DM spike \cite{Gondolo_1999, Sadeghian_2013, Ferrer_2017, Ullio_2001, bertone2024darkmattermoundsrealistic} at the Galactic Center is remarkably resilient to gravitational perturbations from both stellar and BH populations. Accounting for stellar scattering from the nuclear star cluster and the influence of the S2 star, we show that the density remains essentially intact in the innermost regions relevant for gravitational-wave dephasing ($10^{-5}~\mathrm{pc}$). We further find that successive mergers with a population of $\sim 10~\mathrm{M}_\odot$ BHs over $10~\mathrm{Gyr}$ only modestly reduce the density and induce a slight steepening of the spike. These results suggest that DM spikes around galactic centers can survive over cosmological timescales, preserving potentially observable signatures in gravitational-wave systems despite long-term perturbations from stars and stellar-mass compact objects.\\

\section*{Acknowledgments}
We thank Enrico Barausse and the participants in the IFPU's focus week ``Gravitational Wave Probes of Black Hole Environments'' for stimulating discussions when this idea was initiated. We also thank Luca Broggi for valuable discussions on the Fokker-Planck equation, and Rodrigo Vicente and Matteo Sadun Bordoni for their feedback and insight. TK and GB gratefully acknowledge the support of the Dutch Research Council (NWO) through an Open Competition Domain Science-M grant, project number OCENW.M.21.375. FS and BJK acknowledge funding from the \textit{Consolidaci\'on Investigadora} Project \textsc{DarkSpikesGW}, reference CNS2023-144071, financed by MCIN/AEI/10.13039/501100011033 and by the European Union ``NextGenerationEU"/PRTR.

\appendix

\section{Diffusion coefficients} \label{app:coeffs}

The diffusion coefficients appearing in \cref{sec:nsc} are related to the average rate of change of a particles energy, or normalized angular momentum, $\langle \Delta x \rangle_t$ due to gravitational encounters with stars, through the equations
\begin{align}
    D_\calE &= - \langle \Delta \calE \rangle_t + \frac{1}{2} \frac{\partial \langle (\Delta \calE)^2 \rangle_t}{\partial \calE} + \frac{1}{2} \frac{\partial \langle \Delta \calE \Delta \mathcal{R} \rangle_t}{\partial \mathcal{R}}\,,\\
    D_\mathcal{R} &= - \langle \Delta \mathcal{R} \rangle_t + \frac{1}{2} \frac{\partial \langle (\Delta \mathcal{R})^2 \rangle_t}{\partial \calE} \!+ \frac{1}{2} \frac{\partial \langle \Delta \calE \Delta \mathcal{R} \rangle_t}{\partial \calE},\\
    D_{\calE\calE} &= \frac{\langle (\Delta \calE)^2 \rangle_t}{2}\,, \qquad D_{\mathcal{R}\mathcal{R}} = \frac{\langle (\Delta \mathcal{R})^2 \rangle_t}{2}\,,\\
    D_{\calE\mathcal{R}} &= \frac{\langle \Delta\calE \Delta \mathcal{R}\rangle_t}{2}\,.
\end{align}

The instantaneous changes in the scaled angular momentum $\mathcal{R} = 2\calE h^2/(GM)^2 \in (0, 1]$; can themselves be related to those changes in energy and angular momentum simply by differentiating the definition of $\mathcal{R}$ to obtain,
\begin{align}
    (GM)^2 \langle \Delta \mathcal{R}\rangle &= 4\calE h \langle \Delta h\rangle +2h^2 \langle \Delta \calE\rangle \\
    &\quad+ 2\calE \langle (\Delta h)^2\rangle +4h\langle \Delta \calE \Delta h\rangle\,,\nonumber\\
    (GM)^4 \langle (\Delta \mathcal{R})^2\rangle &= 16\calE^2 h^2 \langle (\Delta h)^2\rangle +4h^4 \langle (\Delta \calE)^2\rangle \\
    &\quad+ 16\calE h^3 \langle \Delta \calE \Delta h\rangle \,,\nonumber\\
    (G M)^2 \langle \Delta \calE \Delta \mathcal{R}\rangle &= 4 \calE h \langle \Delta \calE \Delta h\rangle + 2h^2 \langle (\Delta \calE)^2\rangle \,.
\end{align}

\section{Gravitational Encounters} \label{app:encounters}

In this section we derive the average  rate of the first and second moments of energy and angular momentum of a DM particle due to gravitational scattering with stars. While we follow the derivation found in \cite{2013degn.book.....M}, we generalize it to capture the relative velocity kinematics between particles of the two populations.

Let a DM particle's orbit be described by the vector $\mathbf{r} = r\,(\cos\theta, \sin\theta, 0)$ lying in the x-y plane, and therefore has a velocity $\mathbf{v} = \dot{\mathbf{r}} = v \, r\,(\cos\phi_v, \sin\phi_v, 0)$, where $\theta$ is the true anomaly and $\phi_v$ is the angle between the velocity and the x-axis. Additionally, let a star encounter the particle with velocity $\mathbf{v}_s$ sampled from the stellar velocity distribution $f(\mathbf{v}_s)$. It is convenient to express $\mathbf{v}_s$ in terms of its orientation with respect to $\mathbf{v}$, meaning we can write $\mathbf{v}_s = v_s (\sin\xi \cos\psi \,\bm{\hat{\theta}}_v +\sin\xi \sin\psi \, \bm{\hat{\phi}}_v +\cos\xi \,\mathbf{\hat{v}})$, where $\cos\xi \equiv \mathbf{\hat{v}}_s \cdot \mathbf{\hat{v}}$ and $\psi$ a trivial azimuthal angle in the setup, and $\bm{\hat{\theta}}_v = -\mathbf{\hat{z}}$, $\bm{\hat{\phi}}_v = -\sin\phi_v \, \mathbf{\hat{x}} +\cos\phi_v \, \mathbf{\hat{y}}$ are vectors perpendicular to $\mathbf{v}$.

A single gravitational encounter between a DM particle and a star with impact parameter $b$ and relative velocity $\mathbf{V}_0 = \mathbf{v}_s-\mathbf{v}$ will induce a change to the velocity of the particle, $\delta \mathbf{v}$, which can be decomposed to a parallel and perpendicular part, $\delta \mathbf{v} = \delta \mathbf{v}_\parallel + \delta \mathbf{v}_\perp$ to the initial relative velocity vector \cite{binney}, where the magnitudes are given by
\begin{align}
    \delta v_\parallel = \frac{m_s}{m_\chi +m_s}\frac{2V_0}{1 +b^2/b_{90}^2}\,,\quad
    \delta v_\perp = \frac{b}{b_{90}} \delta v_\parallel\,.
\end{align}
and $b_{90} = G(m_s+m_\chi)/V_0^2$. The perpendicular component is parallel to the impact parameter vector $\mathbf{b} = b\, ( \bm{\hat{\theta}}_0 \sin \chi - \bm{\hat{\phi}}_0 \cos\chi)$, where $\bm{\hat{\theta}}_0 = (\cos\theta_0\cos\phi_0, \cos\theta_0\sin\phi_0, -\sin\theta_0)$, and $\bm{\hat{\phi}}_0 = (-\sin\phi_0, \cos\phi_0, 0)$ form an orthonormal basis alongside $\mathbf{\hat{V}}_0 = (\sin\theta_0\cos\phi_0, \sin\theta_0\sin\phi_0, \cos\theta_0)$. These angles can be related with those of $\mathbf{v}$ and $\mathbf{v}_s$ through the trigonometric relations
\begin{align}
    \sin\theta_0 \cos\phi_0 &= \frac{-v_s\sin\xi\sin\psi\sin\phi_v +(v_s\cos\xi-v)\cos\phi_v}{V_0}\,,\nonumber\\
    \sin\theta_0\sin\phi_0 &= \frac{v_s\sin\xi\sin\psi\cos\phi_v +(v_s\cos\xi-v)\sin\phi_v}{V_0}\,,\nonumber\\
    \cos\theta_0 &= -\frac{v_s\sin\xi\cos\psi}{V_0}\,.
\end{align}

To calculate the diffusion coefficients we start by observing the change in energy $\calE$ and angular momenta $h$, $2\delta \calE = v^2 -|\mathbf{v} +\delta\mathbf{v}|^2$, and $h \delta h \approx \mathbf{h} \cdot \mathbf{r} \times \delta \mathbf{v}$. The diffusion coefficients correspond to the total change in the appropriate parameter, e.g. $x$, during a particle's orbit whereby each time interval $\Delta t$ the change is attributed to interacting with a rate of stars following a velocity distribution in the particle frame and a cross-section of particles, meaning
\begin{equation} \label{eq:velocity_distribution_integration}
    \langle \Delta x \rangle = n_s(r) \int \delta x \,b \, \mathrm{d}b\,\mathrm{d}\chi \, V_0\, \Delta t\, f_s(\mathbf{V}_0, m_s)\, \mathrm{d}^3 \mathbf{V}_0\, \mathrm{d}m_s\,,
\end{equation}
and
\begin{equation}
    \langle \Delta x \rangle_t = \frac{1}{T} \int_0^T \frac{\langle \Delta x\rangle }{\Delta t} \, \mathrm{d}t = \frac{2}{hT} \int_0^\pi r^2 \frac{\langle \Delta x\rangle}{\Delta t}  \,\mathrm{d}\theta \,,
\end{equation}
where $T = \sqrt{2} \pi GM\calE^{-3/2}/2$ is the orbital period.

Integrating over $\chi$ will simplify several terms in the changes, leaving the following surviving contributions up to $\mathcal{O}(\delta v^3)$:
\begin{align}
    \delta \calE &= - \mathbf{v} \cdot \delta \mathbf{v}_\parallel -\frac{\delta v^2}{2} \,,\\
    (\delta\calE)^2 &= |\mathbf{v} \cdot \delta \mathbf{v}_\parallel|^2
    + |\mathbf{v} \cdot \delta \mathbf{v}_\perp|^2\,,\\
    h \delta h &= \mathbf{h}\cdot \mathbf{r} \times \mathbf{\delta v_\parallel}\,,\\
    h^2 \! \left(\delta h\right)^2 &= |\mathbf{h}\cdot \mathbf{r} \times \delta\mathbf{v}_\parallel|^2 + |\mathbf{h}\cdot \mathbf{r} \times \delta \mathbf{v}_\perp|^2\,,\\
    h \delta \calE \delta h &= - (\mathbf{v}\cdot\delta \mathbf{v}_\parallel)(\mathbf{h}\cdot \mathbf{r}\times \delta \mathbf{v}_\parallel) \\
    &\quad-(\mathbf{v}\cdot\delta\mathbf{v}_\perp)(\mathbf{h} \cdot \mathbf{r} \times \delta\mathbf{v}_\perp)\,.\nonumber
\end{align}

Assuming isotropy in the stellar distribution to simplify our calculation, we eventually derive:
\begin{widetext}
\begin{align}
    \frac{\langle \Delta \calE \rangle}{\Delta t} &= -\frac{4\pi G^2 n_s}{v} \langle m_s^2 \ln\Lambda \rangle \left[ E_1 -m_\chi \frac{\langle m_s \ln\Lambda \rangle}{\langle m_s^2 \ln\Lambda \rangle} F_0 \right], \qquad\!\! 
    \frac{\langle \Delta h\rangle}{\Delta t} = -\frac{4\pi G^2 n_s h}{v^3} F_0 \langle m_s^2 \ln\Lambda \rangle \left[ 1 + m_\chi \frac{\langle m_s \ln\Lambda \rangle}{\langle m_s^2 \ln\Lambda \rangle} \right],\nonumber\\
    \frac{\langle(\Delta \calE)^2\rangle}{\Delta t} &= \frac{4\pi G^2 n_s v}{3} \big[ 2\langle m_s^2 \ln\Lambda \rangle \left(E_1 +F_2\right) +3\langle \lambda_2 m_s^2 \rangle \left( F_0 -F_2\right)\big]\,,\qquad \qquad
    \frac{\langle \Delta \calE \Delta h \rangle}{\Delta t} =- \frac{h}{v^2} \frac{\langle(\Delta \calE)^2\rangle}{\Delta t} \,,\\
    \frac{\langle(\Delta h)^2\rangle}{\Delta t} &= \frac{2\pi G^2 n_s h^2}{v^3} \left[ 2 \langle m_s^2 \ln\Lambda \rangle \left( F_0 \frac{r'^2}{r^2} +\frac{F_2}{3}\left(2-\frac{r'^2}{r^2}\right)  +\frac{2E_1}{3} \left(1 +\frac{r'^2}{r^2}\right)\right) +\langle \lambda_2 m_s^2\rangle \left(F_0 -F_2\right) \left(2-\frac{r'^2}{r^2}\right)\right]\,.\nonumber
\end{align}
\end{widetext}
where $\Gamma = 2\pi G^2 m_s^2 \,n_s$, $r'$ is the partial derivative of the particle's position with respect to $\theta$, the brackets $\langle \cdots \rangle$ denote an average over the stellar mass spectrum, and
\begin{align}
    E_1(x) \equiv v \int_v^\infty \!\!f_s\,\mathrm{d}\ln v_s &= \frac{2\,x}{\sqrt{\pi}} \frac{\Gamma(\gamma_s+1)}{\Gamma(\gamma_s+1/2)} \left(1-x^2\right)^{\gamma_s-1/2}\!\!, \nonumber \\
    F_0(x) \equiv \int_0^v f_s\,\mathrm{d} v_s &= \mathcal{I}\left(x^2; 3/2, \gamma_s-1/2\right), \\
    F_2(x) \equiv v^{-2} \int_0^v v_s^2 \, f_s\,\mathrm{d} v_s &=\frac{3\mathcal{I}\left(x^2; 5/2, \gamma_s-1/2\right)}{2x^2(1+\gamma_s)}\,, \nonumber
\end{align}
are skewed velocity moments of the stellar distribution, e.g. $F_0$ represents the fraction of stars moving more slowly compared to the DM particle in question\footnote{In this work we assume the stellar distribution to be fixed, and therefore we can take the analytical form of these functions. One however, may find them numerically in a time-dependent system by adopting the usual marginalization techniques for $\mathcal{R}$ \cite{2013degn.book.....M}}. Also, $\mathcal{I}$ is the normalized incomplete beta function, and \linebreak $x = v/v_\mathrm{max}(r)$. Finally, we have
\begin{equation}
    \ln\Lambda = \ln\sqrt{\frac{b_\mathrm{max}^2 +b_{90}^2}{b_\mathrm{min}^2 +b_{90}^2}} \approx \ln\frac{M}{m_s} +\ln\left( 2 - \frac{r}{a}\right)\,,
\end{equation}
and
\begin{equation}
    \lambda_2 = \frac{\Lambda^2 -1}{\Lambda^2(1 +b^2_\mathrm{min}/b_{90}^2)} \approx 1 \,,
\end{equation}
where $a$ is the particle's semi-major axis, and we've taken $b_\mathrm{min}/b_\mathrm{90} \ll 1$. For a stellar population, $b_\mathrm{min}$ can be naturally identified with the stellar radius, since particle deflection within the interior is strongly suppressed. Consequently, extended and less compact objects, such as red giants, are expected to contribute negligibly to the dynamical-friction back-reaction of the medium \cite{karydas2025measuringneutronstarequation}. A quantitative assessment of this effect is deferred to future work.

In the conventional derivation of the Fokker-Planck equation, one finds a different, simplified dependence on the velocity moment functions for these energy and angular momentum terms. Notably, a direct consequence of modeling the relative velocity between test (DM) and field (star) particle velocities in our derivation, is that in the limit $m_\mathrm{\chi} \ll m_s$, there is a non-zero and non-small drift in energy, which would have vanished identically otherwise because of the assumptions $V_0 = v$.

\section{Impact of the Stirring Effect} \label{app:stirring}

To quantify the impact of the three-body stirring effect, we study the evolution of the system using a Monte Carlo approach in phase space, based on stochastic differential equations rather than master equations, in analogy with the method used to solve the Fokker–Planck equation in \cref{sec:nsc}. A simple version of this technique was describe in Ref.~\cite{Kavanagh:2024lgq} to model the close and far scattering of particles. We provide a brief summary of our more detailed approach below, with full details to appear in an upcoming publication~\cite{LongTermDepletion}. For simplicity, we take the inspirals to be quasi-circular which maximizes the depletion \cite{karydas2024sharpeningdarkmattersignature}.

In the Monte Carlo approach, each particle is described by its energy $\mathcal{E}$, total angular momentum $h$ and the $z$-component of the angular momentum $h_z$. 
For each orbit of the binary, we draw a position $\mathbf{r}$ and velocity $\mathbf{v}$ for each particle, weighted by the time spent at that position, given $(\mathcal{E}, h, h_z)$. We then compute the perpendicular distance $b$ of the DM particle from the orbit of the companion. If $b < b_\mathrm{max}$, then the particle experiences a close encounter, receiving a velocity kick which depends on $b$ and $\mathbf{v}$.

Additionally, we incorporate the effect of 3-body stirring. This contribution arises from the explicit time dependence of the Hamiltonian associated with the orbiting companion, in which case strictly conserved quantities no longer exist. For particles which do not experience a close encounter with a companion at a separation $a$ from the primary, we update the particle's velocity according to:
\begin{equation}
    \Delta v \propto \frac{Gm a^2 \, T_\mathrm{orb}}{r^4}\,.
\end{equation}
This shift in velocity may increase and decrease the particle's energy, with the direction and precise magnitude of $\Delta \mathbf{v}$ depending on the relative orbital orientation between the particle and the companion (c.f.~Ref.~\cite[Eq. C10]{Kavanagh:2024lgq}).

From these velocity kicks, we update the integrals of motion $(\mathcal{E}, h, h_z)$ of each particle. Particles with $\mathcal{E} < 0$ are considered unbound from the system, and the spherically-symmetric density profile can be reconstructed from the $(\mathcal{E}, h)$ values of the bound particles. This approach provides good agreement with the density profiles observed in $N$-body simulations over tens of thousands of orbits of the binary~\cite{LongTermDepletion}.

\begin{figure}[ht!]
    \centering
    \hspace*{-0.08\columnwidth}
    \includegraphics[width=0.95\linewidth]{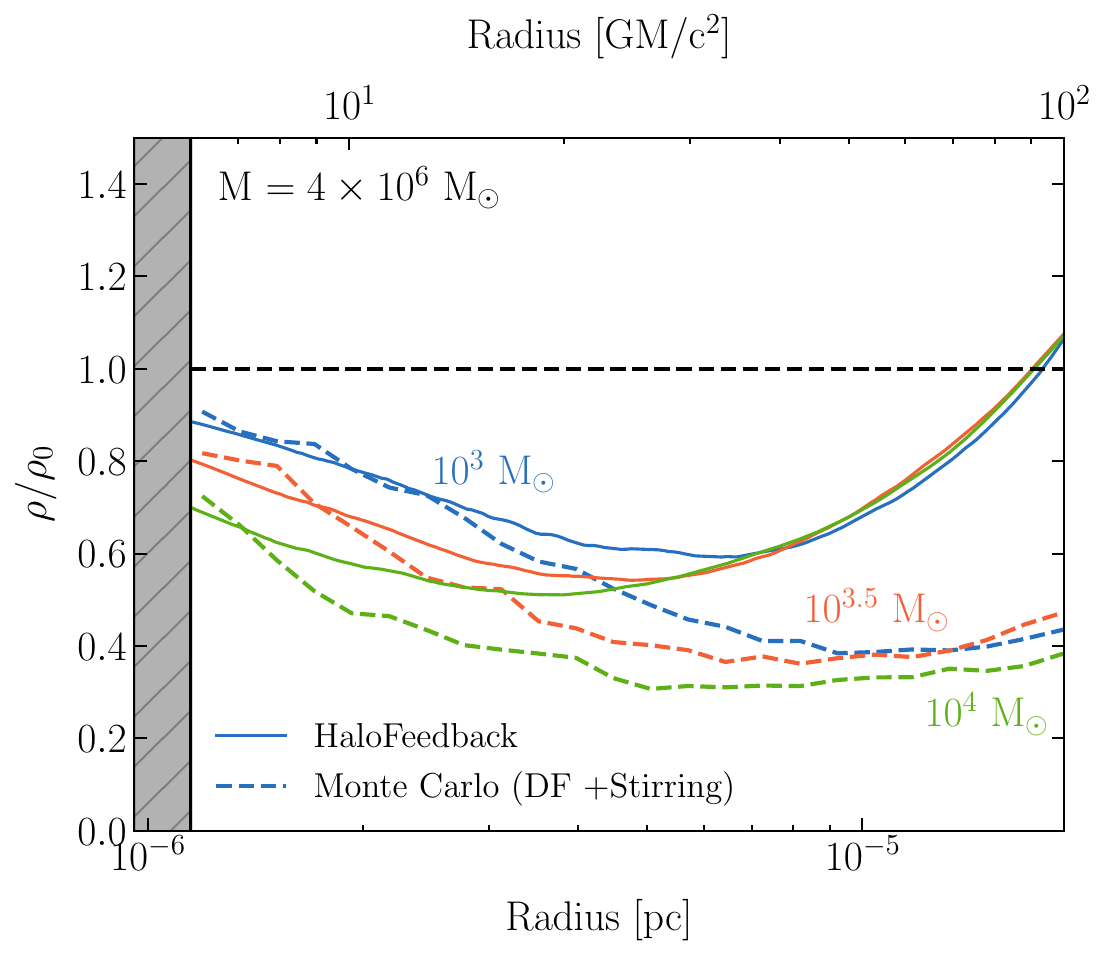}
    \caption{\textbf{Impact of the stirring effect on the dark matter density profile during inspiral.} The companion is initialised at a radius of $100\,GM/c^2 \approx 1.9 \times 10^{-5}\,\mathrm{pc}$. \label{fig:stirring}}
\end{figure}

In practice, we evaluate this effect by tracking the evolution of the DM distribution in numerical simulations of inspirals with varying companion masses, using the standard \texttt{HaloFeedback} approach described in the main text and using the MC approach, which includes the stirring contribution. In particular, we consider companions with $m = \left\{ 10^4, 3.2 \times 10^3, 10^3\right\}M_\odot$ while keeping the mass of the primary fixed at \linebreak $M = 4\times10^6~M_\odot$. Extending this analysis to lower companion masses, such as $m = 10~M_\odot$ which would more closely reflect the EMRI population studied in \cref{sec:emris}, is computationally prohibitive. This is because the present approach requires resolving the evolution on an orbit-by-orbit basis, with the number of orbits scaling as $N_\mathrm{orb} \propto 1/q$, making low-mass inspirals significantly more expensive to simulate. Nonetheless, we observe that the impact of the stirring contribution decreases with decreasing mass ratio; the behaviour observed in the simulated cases already allows us to reliably infer the expected impact in this lower-mass regime, and no additional simulations are required.

The results are shown in \cref{fig:stirring}. We find that the stirring effect significantly modifies the density profile in the outer regions of the spike, where the cumulative impact of the time-dependent potential is most pronounced. In contrast, the innermost regions (within the break radius~\cite{Coogan_2022}) are well described by strong scattering alone, with only a subdominant contribution from stirring. In this regime, the timescale for inspiral due to gravitational waves is shorter than the typical timescale for depletion, with the stirring effect becoming more relevant on longer timescales~\cite{Kavanagh:2024lgq}. This indicates that neglecting the stirring effect provides a good approximation when focusing on the innermost density profile.

Finally, we comment on the emergence of spherical asymmetry after the completion of an inspiral. In our heaviest simulation ($m=10^4~\mathrm{M}_\odot$), the DM distribution exhibits substantial departures from spherical symmetry at large radii, driven by the preferential depletion of near-polar DM orbits as opposed to planar orbits whose co-rotation with the companion can suppress their scattering rate. Towards the center, however, the system regains spherical symmetry: at $r=20GM/c^2$ the relative density variation between polar and equatorial orientations is $\sim 1-2$, while at $r = 9GM/c^2$ we detect no deviation from spherical symmetry.

\bibliography{bibliography}

\end{document}